\documentclass[%
 reprint,showkeys,
 amsmath,amssymb,
 aps,
pra,
]{revtex4-2}

\usepackage{graphicx}% Include figure files
\usepackage{float}
\usepackage{dcolumn}% Align table columns on decimal point
\usepackage{bm}% bold math
\usepackage{amsmath}
\usepackage{booktabs}
\usepackage{xcolor}

\newcommand{\e}[1]{\text{#1}}

\begin{document}

\preprint{APS/123-QED}

\title{Microwave spectroscopy assisted by electromagnetically  induced transparency near natural Förster resonance on Rubidium}% Force line breaks with \\

\author{Naomy Duarte Gomes$^{1}$, Daniel Varela Magalhães$^{1}$, Jorge Douglas Massayuki Kondo$^{1,2}$, Luis Gustavo Marcassa$^{1}$}

\affiliation{$^{1}$Instituto de F\'{i}sica de S\~{a}o Carlos, Universidade de S\~{a}o Paulo, Caixa Postal 369, 13560-970, S\~{a}o Carlos, SP, Brasil.}
\affiliation{$^{2}$Departamento de F\'{i}sica, Universidade Federal de Santa Catarina, Florianópolis 88040-900, SC, Brasil}

\date{\today}% It is always \today, today,

\begin{abstract}

In this work, we precisely measure the transition energies between the Rydberg state n$\text{D}_{5/2}$ to the nearby Rydberg states $(\e{n}+2)\text{P}_{3/2}$, $(\e{n}-2)\text{F}_{7/2}$ for the range $41\leq \e{n}\leq46$. This was done by carrying out microwave spectroscopy via Electromagnetically Induced Transparency (EIT) in a room temperature vapor reference cell of Rubidium, which is similar to the experimental approach followed by Li et al. [Results in Physics \textbf{29}, 104728 (2021)]. This range is interesting because there is a quasi Förster resonance between the atomic pair $43\text{D}_{5/2}+43\text{D}_{5/2}$ and $45\text{P}_{3/2}+41\text{F}_{7/2}$. We compared the obtained results with numerically calculated transition energies based on previously tabulated quantum defect numbers by various research groups using both hot and ultra-cold atomic samples. Our data are more consistent with measurements made within ultra-cold atomic systems [Phys. Rev. A \textbf{67}, 052502 (2003), Phys. Rev. A \textbf{74}, 054502 (2006)].
\end{abstract}

%Use showkeys class option if keyword
\maketitle

%\tableofcontents

\section{\label{sec:introduction}INTRODUCTION}

In recent years, the exploration of the unique properties of highly excited Rydberg atoms has become pivotal for advancing several quantum technologies, spanning from quantum information \cite{RevModPhys.82.2313,Liu:18}, computation \cite{PRXQuantum.2.030322,10.1093/nsr/nwy088}, simulation \cite{Nature_browaeys,Weimer2010}, highly sensitive electromagnetic sensors \cite{sedlacek2012microwave,holloway2017electric,holloway2014broadband,holloway2014sub,fan2015atom,PhysRevLett.111.063001}, to understanding the physical background behind non-equilibrium phase transitions in hot atomic vapors \cite{PhysRevLett.111.113901,PhysRevLett.131.143002}. These inquiries also extend to the characterization of many fundamental atomic properties, such as atomic lifetime \cite{holloway2014broadband}, and quadrupolar polarizability information of the alkali ionic core \cite{atoms4030021,PhysRevA.102.062817,PhysRevA.102.062818}. In order to explore the highly precision state energies and transition energies embedded in Rydberg atomic states, fundamental information is imperative, specifically their intrinsic transition dipole moments ($\mu_{d}$), which play a central role in electromagnetic sensing and entanglement protocols based on Rydberg-Rydberg interactions \cite{PhysRevLett.104.010502,PhysRevLett.102.240502,Urban2009}. The calculation of these precise energies of Rydberg transitions is paramount in predicting accurately the behavior of atomic or molecular samples under study.

Determination of the Rydberg energy spectrum in alkali atomic samples involves an effective quantum number, denoted $\e{n}^* = \e{n} - \delta_{\e{n},\e{l},\e{j}}$, where the deviation $\delta_{\e{n}, \e{l},\e{j}}$, is known as quantum defect \cite{gallagher2006rydberg}, and is experimentally obtained. Quantum defects exhibits a dependence on the orbital and total angular momentum (l, j) and a slight dependence on the principal quantum number $\e{n}$. Numerous research groups have systematically measured quantum defects for various alkali atomic samples, as in the works of Wenhui Li et al. \cite{PhysRevA.67.052502} and Jianing Han et al. \cite{PhysRevA.74.054502} providing a comprehensive set for ${}^{85}\text{Rb}$. They meticulously tabulated quantum defects for $\e{n}S_{1/2}$, $\e{n}\e{P}_{1/2,3/2}$, $\e{n}\e{D}_{3/2,5/2}$, $\e{n}\e{F}_{5/2,7/2}$, using an ultra-cold sample held within a magneto-optical trap. Other research groups, such as Sanguinetti et al. \cite{Sanguinetti_2009} and Johnson et al. \cite{Johnson_2010}, have explored identical Rydberg states, revealing slightly different quantum defects by measuring $\e{n}\e{P}_{3/2}$ and $\e{n}\e{F}_{7/2}$ using vapor cell spectroscopy. Shaohua Li et al. \cite{LI2021104728} also investigated the same Rydberg states, but using microwave (MW) assisted Electromagnetic Induced Transparency (EIT), starting from the $\e{nD}_{5/2}$ state. 

A pair of $nD_{5/2}$ Rydberg atoms contains a natural Förster resonance where the atomic pairwise energy $43\e{D}_{5/2}+43\e{D}_{5/2}$ has an almost degenerate energy with the $45\e{P}_{3/2}+41\e{F}_{7/2}$ atomic pair, leading to strong dipole Rydberg-Rydberg two-body interactions \cite{PhysRevA.90.023413}. This is analogous to Förster Resonance Energy Transfer (FRET), which is a mechanism of energy transfer between two light-sensitive molecules without photon emission. It has been proposed that such a process may be used for the implementation of quantum gates, allowing precise control of the qubit \cite{PhysRevA.92.042710}.

In this work, we present a spectroscopic analysis built up via EIT using two optical photons and one microwave photon to obtain the energy spectrum of Rydberg states. Specifically, we explore the states $(\e{n}+2)\e{P}_{3/2}$ and $(\e{n}-2)\e{F}_{7/2}$, which can be coupled with the initial target Rydberg state $\e{nD}_{5/2}$ for $41\leq \e{n}\leq46$. We carried directly measurements of energy differences for a natural Förster resonance at n=43, which are compared with theoretical values using quantum defects already tabulated in the literature \cite{PhysRevA.67.052502, PhysRevA.74.054502,Sanguinetti_2009,Johnson_2010,LI2021104728}. Our findings show closer alignment with measurements conducted using ultra-cold atomic samples \cite{PhysRevA.67.052502, PhysRevA.74.054502}.

\section{\label{sec:experimental}EXPERIMENTAL SETUP and METHODS}

\begin{figure}[ht]
    \centering
    \includegraphics[scale=0.43]{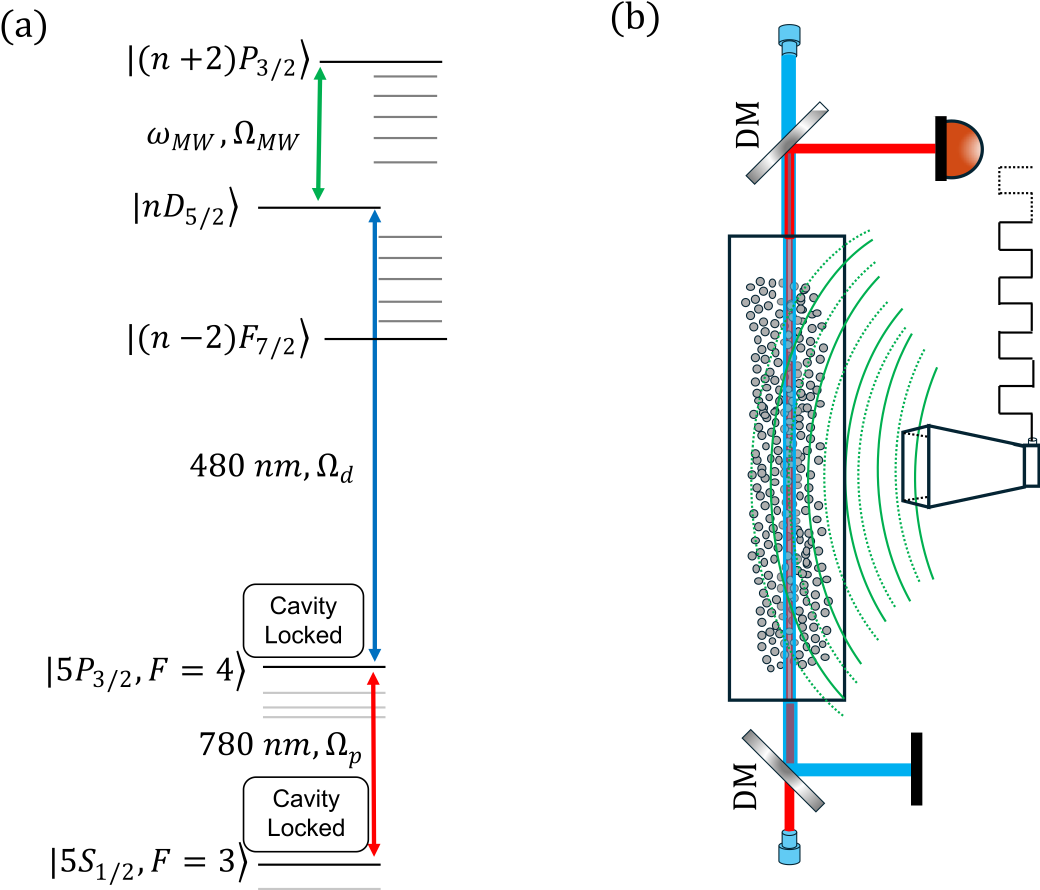}% Here is how to import EPS art
    \caption{\label{fig:epsart} (a) Energy level diagram for MW spectroscopy using EIT, (b) $\text{Rb}$ reference cell, horn microwave antenna.}
\end{figure}

The experimental setup utilized in this study follows the design described in \cite{duarte2022polarization}. Clearly, adjustments and modifications were introduced, tailoring the setup to the unique requirements of our research questions and enhancing experimental precision, as shown in Figure \ref{fig:epsart}. In brief, we used a 75 mm Rb vapor reference cell at room temperature. A diode laser at 780 nm couples the ground state to the intermediate state $5\e{S}_{1/2},(\e{F}=3) \rightarrow 5\e{P}_{3/2},(\e{F'}=4)$; this is called the probe beam and operates in the weak regime when its intensity is much lower than the transition saturation intensity. A second laser operating at 480 nm couples the intermediate state to the first initial target Rydberg state $5\e{P}_{3/2},(\e{F}=4) \rightarrow {\e{nD}_{5/2}}$ \cite{PhysRevA.93.012703} and is called the coupling beam. Both optical beams are frequency locked at zero detuning, using a thermally stabilized home-made Fabry-Perot optical cavity \cite{RodriguezFernandez2023}, with laser linewidths of 100 kHz. The microwave photon is generated by an eightfold circuit consisting of a passive doubler (Model ZX90-2-24-S+ from Minicircuits) and quadrupler (Model 934VF-10/385 from Mi-Wave), producing frequencies in the range of 50 to 75 GHz. This circuit connects the initial Rydberg target state to nearby Rydberg states $\e{nD}_{5/2} \rightarrow (\e{n}+2)\e{P}_{3/2}$ and $(\e{n}-2)\e{F}_{7/2}$ through a dipole one-photon transition. The MW signal is provided by a commercial two-channel MW generator ( Windfreak Technologies, model SynthPRO). To improve the signal-to-noise ratio, the probe beam optical signal is processed by a lock-in amplifier at 1kHz modulation of the microwave field. A round horn antenna (Model WR15 UG-385/U-Pasternack) feeds the MW to the atoms and is positioned at a distance of 40 cm from the experimental cell to obtain quasi-plane wave microwave radiation. 

\begin{figure}[ht]
    \centering
    \includegraphics[scale=0.6]{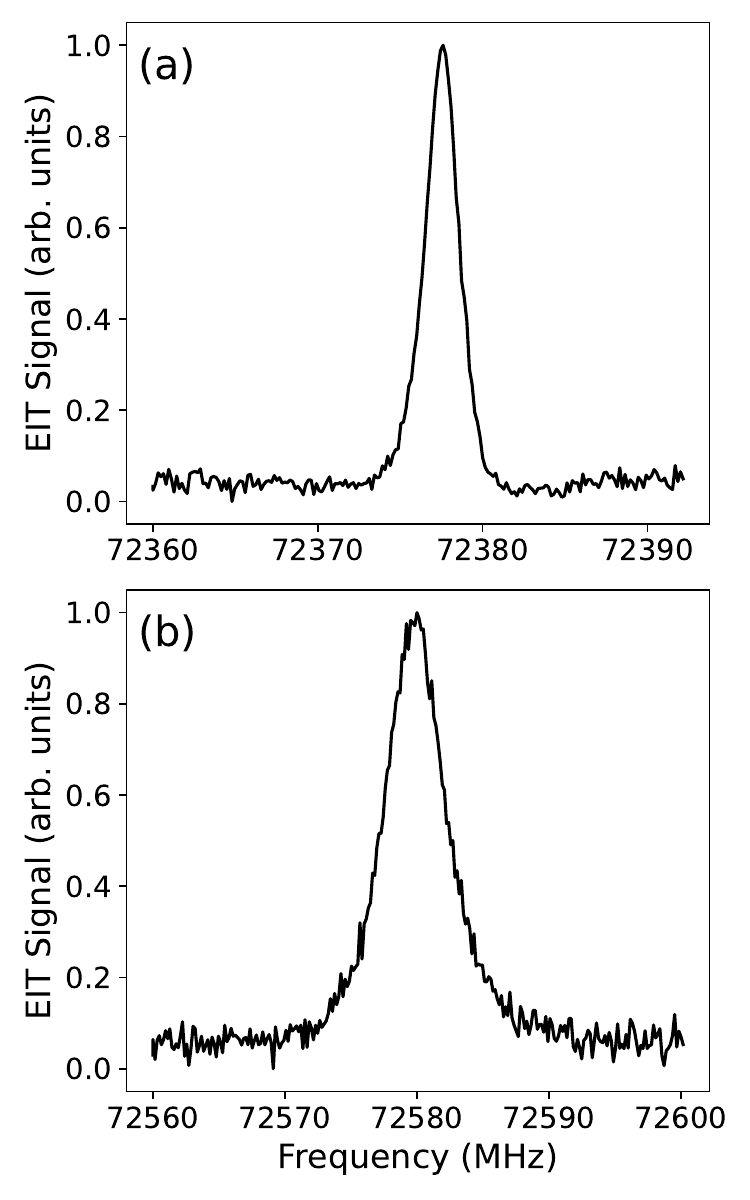}
    \caption{\label{fig:epsart3} EIT typical spectra: (a) $41D_{5/2} \rightarrow 43P_{3/2}$, (b) $41D_{5/2} \rightarrow 39F_{7/2}$.}
\end{figure}

The optical beams counter-propagate and are all linearly polarized, overlapping at the center of the cell with a $180~\mu$m waist. The probe laser Rabi frequency is $\Omega_{p}=2\pi \times 2.32~\text{MHz}$ and is constantly monitored by a photodiode. The coupling laser has $\Omega_{d}=2\pi \times3.56$ MHz. The MW photon Rabi frequency is $\Omega_{MW}=2\pi\times0.5$ MHz and is orthogonally aligned in respect to the vapor cell, probe, and coupling beams. The MW power was calibrated using the Autler-Townes splitting for each transition. Figure \ref{fig:epsart3} shows experimental EIT spectra for the $43\e{P}_{3/2}$ and $39\e{F}_{7/2}$ coupled to the initial Rydberg target state $41\e{D}_{5/2}$.

\section{\label{sec:discussion}EXPERIMENTAL ANALYSIS AND DISCUSSION}

We have measured the following transition frequencies between the initial Rydberg target state $\e{nD}_{5/2}$ and the final states for the range of principal quantum numbers $41 \leq \e{n} \leq 46$ (see equations \ref{eq:one} and \ref{eq:two}). The main sources of error for the transition frequency measurements are the frequency positions of the EIT peaks and the instabilities in the lock of the probe and coupling lasers. Both lasers are frequency-stabilized using an optical cavity, with a frequency instability of 100 kHz. The optical cavity itself exhibits a 50 kHz drift per hour. Based on multiple measurements, the standard deviation is 1 MHz for the P and F Rydberg states in the determination of the peak positions. This standard deviation dominates over other errors and will be considered as the maximum error in this work.

\begin{align}
\e{nD}_{5/2} & \rightarrow (\e{n}+2)\e{P}_{3/2}. \label{eq:one} \\
\e{nD}_{5/2} & \rightarrow (\e{n}-2)\e{F}_{7/2}. \label{eq:two}
%\e{nD}_{5/2} & \rightarrow (\e{n}-2)\e{G}. \label{eq:three}\\
%\e{nD}_{5/2} & \rightarrow (\e{n}-2)\e{F} \rightarrow %
%(\e{n}-2)\e{H}.\label{eq:four}
\end{align}

The experimental transition frequencies can be compared to calculated ones, $\nu_{\e{nn'}}$, using Equation \ref{eq:rydbergdel}, where n and n' represent the initial target Rydberg state, $\e{nD}_{5/2}$, and any final state, respectively.

\begin{align}
\nu_{\e{nn'}}=\e{R}_{\e{y}}\e{c}\left(\dfrac{1}{(\e{n}-\delta_{\e{n},\e{l},\e{j}})^2}-\dfrac{1}{(\e{n'}-\delta_{\e{n'},\e{l'},\e{j'}})^2}\right). \label{eq:rydbergdel} 
\end{align}

\begin{align}
\delta_{\e{n},\e{l},\e{j}}\approx \delta_0+\dfrac{\delta_2}{(\e{n}-\delta_0)^2}. \label{eq:ritz} 
\end{align}

Where $\e{R}_{\e{y}} = 109736.605~\text{cm}^{-1}$ is the Rydberg constant for ${}^{85}\text{Rb}$, $\e{c} = 2.99792458\times10^{10}~\text{cm/s}$ is the speed of light, and $\delta_{\e{n},\e{l},\e{j}}$ is the quantum defect, expressed through the modified Rydberg-Ritz coefficients \cite{Lorenzen_1983, Sansonetti:85, PhysRevA.67.052502} as presented in Equation \ref{eq:ritz}. Here, $\delta_0$ and $\delta_2$ are constant coefficients experimentally tabulated for each orbital angular momentum quantum number l and the total angular momentum j when accessible.
 
\begin{table}[ht]
\caption{\label{tab:table1}%
Transition frequencies from $\e{nD}_{5/2}\rightarrow(\e{n}+2)\e{P}_{3/2}$ measured in this work are compared with the calculated transition frequencies using quantum defects tabulated by W. Li et al. \cite{PhysRevA.67.052502}. The relative deviation is also shown.
}
\begin{ruledtabular}
\begin{tabular}{cccc}
\textrm{n}&
\textrm{$\nu_{nD_{5/2},(n+2)P_{3/2}}$}&
\textrm{$\nu_{nD_{5/2},(n+2)P_{3/2}}$}&
\textrm{Relative Deviation}\\
\textrm{}&
\textrm{$\text{(MHz)}$}&
\textrm{W. Li et al.\cite{PhysRevA.67.052502}}&
\textrm{($\times10^{-5}$)}\\
\colrule
41 & 72377$\pm$ 1 & 72378.38$\pm$ 0.11 & -1.9$\pm$ 1.5\\
42 & 67212$\pm$ 1 & 67213.37$\pm$ 0.11 & -2.0$\pm$ 1.6\\
43 & 62527$\pm$ 1 & 62528.29$\pm$ 0.10 & -2.0$\pm$ 1.7\\
44 & 58267$\pm$ 1 & 58268.68$\pm$ 0.09 & -2.9$\pm$ 1.9\\
45 & 54386$\pm$ 1 & 54387.31$\pm$ 0.08 & -2.4$\pm$ 2.0\\
46 & 50844$\pm$ 1 & 50843.13$\pm$ 0.08 & 1.7$\pm$ 2.1\\
\end{tabular}
\end{ruledtabular}
\end{table}

Initially, we compared our transition frequency measurements for equations \ref{eq:one} and \ref{eq:two}. It is important to note that this comparison can only be done with references \cite{PhysRevA.67.052502,PhysRevA.74.054502}. Table \ref{tab:table1} displays the experimental data alongside calculated values using quantum defects tabulated by W. Li et al. \cite{PhysRevA.67.052502} for the transition $\e{nD}_{5/2}\rightarrow(\e{n}+2)\e{P}_{3/2}$. Table \ref{tab:table2} presents a comparison of relative transition frequencies of $\e{nD}_{5/2}\rightarrow(\e{n}-2)\e{F}_{7/2}$ using quantum defects provided by J. Han et al. \cite{PhysRevA.74.054502}. The relative deviation is at the order of $10^{-5}$, which may be due to the fact that our experiment is performed with hot atoms, introducing broadening to the transition peaks due to the atomic velocity distribution within the sample.

\begin{table}[ht]
\caption{\label{tab:table2}%
Transition frequencies from $\e{nD}_{5/2}\rightarrow(\e{n}-2)\e{F}_{7/2}$ measured in this work are compared with the calculated transition frequencies using quantum defects tabulated by J. Han et al.\cite{PhysRevA.74.054502}. The relative deviation is also shown.
}
\begin{ruledtabular}
\begin{tabular}{cccc}
\textrm{n}&
\textrm{$\nu_{nD_{5/2},(n-2)F_{7/2}}$}&
\textrm{$\nu_{nD_{5/2},(n-2)F_{7/2}}$}&
\textrm{Relative Deviation}\\
\textrm{}&
\textrm{$\text{(MHz)}$}&
\textrm{J. Han et al.\cite{PhysRevA.74.054502}}&
\textrm{($\times10^{-5}$)}\\
\colrule
41 & 72579$\pm$ 1 & 72577.46$\pm$ 0.52 & 2.1$\pm$ 2.1\\
42 & 67310$\pm$ 1 & 67308.01$\pm$ 0.46 & 3.0$\pm$ 2.2\\
43 & 62538$\pm$ 1 & 62536.62$\pm$ 0.41 & 2.2$\pm$ 2.2\\
44 & 58209$\pm$ 1 & 58205.80$\pm$ 0.36 & 5.5$\pm$ 2.3\\
45 & 54269$\pm$ 1 & 54265.85$\pm$ 0.32 & 5.8$\pm$ 2.4\\
46 & 50675$\pm$ 1 & 50673.64$\pm$ 0.29 & 2.7$\pm$ 2.5\\
\end{tabular}
\end{ruledtabular}
\end{table}

\begin{figure}[ht]
    \centering
    \includegraphics[scale=0.5]{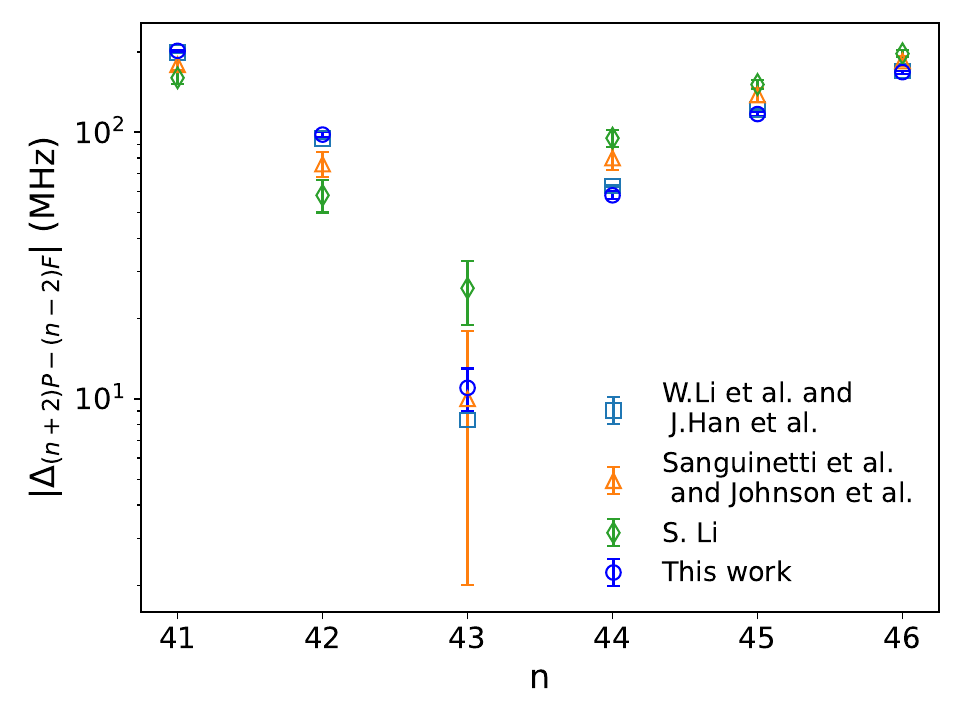}% Here is how to import EPS art
    \caption{\label{fig:epsart4} The energy difference ($|\Delta_{(\e{n}+2)\e{P}-(\e{n}-2)\e{F}}|$) between $(\e{n}+2)\e{P}_{3/2}$ and $(\e{n}-2)\e{F}_{7/2}$ states as a function of n. The data acquired in this work is compared to calculated values using quantum defects coefficients obtained by W. Li and J. Han et al.\cite{PhysRevA.67.052502,PhysRevA.74.054502}, Sanguinetti and Johnson et al. \cite{Sanguinetti_2009,Johnson_2010} and Shaohua Li et al.\cite{LI2021104728}}. 
\end{figure}

We than extract from the experimental transitions frequencies the energy difference between $(\e{n}+2)\e{P}_{3/2}$ and $(\e{n}-2)\e{F}_{7/2}$ Rydberg states and present the absolute value $|\Delta_{(\e{n}+2)\e{P}-(\e{n}-2)\e{F}}|$ (second column in Table \ref{tab:table3}). This parameter was also calculated using the quantum defects obtained experimentally by W. Li and J. Han et al. \cite{PhysRevA.67.052502,PhysRevA.74.054502}, Sanguinetti and Johnson et al. \cite{Sanguinetti_2009,Johnson_2010}, and Shaohua Li et al. \cite{LI2021104728}, and are also presented in Table \ref{tab:table3}. Figure \ref{fig:epsart4} shows $|\Delta_{(n+2)P-(n-2)F}|$ obtained from this work in comparison to references \cite{PhysRevA.67.052502,PhysRevA.74.054502,Sanguinetti_2009,Johnson_2010,LI2021104728}. Particularly, our experimental data are in lines with the values calculated using quantum defects from W. Li and J. Han et al. \cite{PhysRevA.67.052502,PhysRevA.74.054502}, while discernibly mismatching the values from other research groups \cite{Sanguinetti_2009,Johnson_2010,LI2021104728}. A pronounced dependence on the principal quantum number $n$ is evident in this case, due to a naturally occurring Förster resonance for n = 43  \cite{PhysRevA.90.023413,PhysRevA.79.043420}.

\begin{table}[ht]
\caption{\label{tab:table3}%
Energy difference ($|\Delta_{(\e{n}+2)\e{P}-(\e{n}-2)\e{F}}|$) measured in this work is shown below and compared with expected values using quantum defect coefficients from W. Li and J. Han et al. \cite{PhysRevA.67.052502,PhysRevA.74.054502}, Sanguinetti and Johnson et al. \cite{Sanguinetti_2009,Johnson_2010}, and Shaohua Li et al. \cite{LI2021104728}.
}
\begin{ruledtabular}
\begin{tabular}{ccccc}
\textrm{n}&
\textrm{Exp.}&
\textrm{W. Li et al. \cite{PhysRevA.67.052502}}&
\textrm{Sanguinetti et al.\cite{Sanguinetti_2009}}&
\textrm{S. Li\cite{LI2021104728}}\\
\textrm{}&
\textrm{}&
\textrm{J. Han et al. \cite{PhysRevA.74.054502}}&
\textrm{Johnson et al.\cite{Johnson_2010}}&
\textrm{}\\
\textrm{}&
\textrm{(MHz)}&
\textrm{(MHz)}&
\textrm{(MHz)}&
\textrm{(MHz)}\\
\colrule
41 & 202$\pm$ 2 & 199.08$\pm$ 0.63 & 179 $\pm$ 8 & 160 $\pm$ 8\\
42 & 98$\pm$ 2 & 94.65$\pm$ 0.56 & 76 $\pm$ 8 & 58 $\pm$ 8\\
43 & 11$\pm$ 2 & 8.33$\pm$ 0.50 & 10 $\pm$ 8 & 26 $\pm$ 7\\
44 & 58$\pm$ 2 & 62.88$\pm$ 0.45 & 80 $\pm$ 8 & 95 $\pm$ 7\\
45 & 117$\pm$ 2 & 121.47$\pm$ 0.40 & 138 $\pm$ 8 & 151 $\pm$ 6\\
46 & 168$\pm$ 2 & 169.49$\pm$ 0.36 & 185 $\pm$ 8 & 197 $\pm$ 6\\
\end{tabular}
\end{ruledtabular}
\end{table}

\begin{figure}[ht]
    \centering
    \includegraphics[scale=0.49]{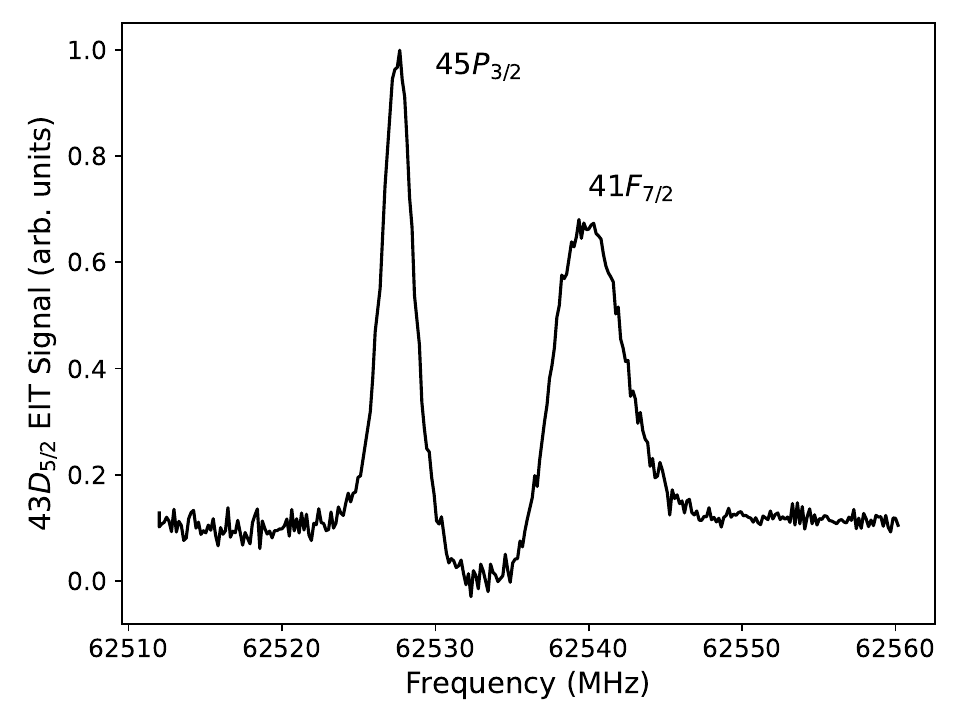}% Here is how to import EPS art
    \caption{\label{fig:epsart5} Transmission signal (EIT) as a function of microwave frequency, coupling $43\e{D}_{5/2}\rightarrow45\e{P}_{3/2}$ and $43\e{D}_{5/2}\rightarrow41\e{F}_{7/2}$ Rydberg state. The measured energy difference from Förster resonance is $11\pm2$ MHz.}
\end{figure}

Figure \ref{fig:epsart5} shows the dependency of the EIT signal on the microwave radiation frequency, driving the $43\e{D}_{5/2}\rightarrow45\e{P}_{3/2}$ and $43\e{D}_{5/2}\rightarrow41\e{F}_{7/2}$ transitions. The measured Förster resonance energy difference is $11\pm2$ MHz, which is twice smaller than predicted by Shaohua Li et al. \cite{LI2021104728} in their available data. Furthermore, there are a few differences between the experimental conditions in the two works. Firstly, Shaohua Li et al. \cite{LI2021104728} performed their measurements using a microwave Rabi frequency 20 times higher than ours, which causes power broadening and shifts the spectral lines. Their atomic transition Q factors ($Q=\nu_{nn'}/ \Delta\nu_{\text{MW}}$) are about 1,000, whereas in our probed range of Rydberg states, it is 5 times larger resulting in higher accuracy. Secondly, they intensity-modulated the coupling laser using a mechanical chopper, limiting the modulation frequency to a few kHz. Furthermore, we observed a clear improvement in the signal-to-noise ratio when modulating the microwave power instead.
 
\section{\label{sec:conclusion}CONCLUSION}

In this study, we employed microwave spectroscopy assisted by EIT to measure the relative transition frequencies for Rydberg atomic energy states of ${}^{85}\text{Rb}$. Specifically, we focus on the states $(\e{n}+2)\e{P}_{3/2}$ and $(\e{n}-2)\e{F}_{7/2}$, all originating from the initial Rydberg target state $\e{nD}_{5/2}$ near a natural occurrence of the Förster resonance. Notably, we identified the lowest energy difference due to the quasi-molecular formation of the excited Rydberg two-body system, pinpointing a natural Förster resonance at n=43, marked by an experimentally measured splitting of $11\pm 2~\text{MHz}$, in excellent agreement with expected values. To validate our experimental results, we compared the relative transition frequencies with those expected, calculated using a modified Rydberg-Ritz equation,  incorporating tabulated quantum defects $\delta_{l,j}$ obtained by various research groups utilizing different methods. Our conclusion is that our results strongly support the quantum defect values generated by Wenhui Li et al. \cite{PhysRevA.67.052502} and Jianing Han et al. \cite{PhysRevA.74.054502} using cold atomic samples. Recent quantum defect measurement in Cs suggests that better results can only be obtained using cold atomic samples and laser sources calibrated against a frequency comb locked  to an atomic clock \cite{Jim}.

\begin{acknowledgments}
This work is supported by grants 2019/10971-0 and 2021/06371-7, S\~{a}o Paulo Research Foundation (FAPESP), and CNPq (305257/2022-6). It was supported by Army Research Office - Grant Number W911NF-21-1-0211.
\end{acknowledgments}

\newpage
\bibliography{main.bib}% Produces the bibliography via BibTeX.

\end{document}